\documentclass[aps,pra,twocolumn,groupaddress]{revtex4-2}
\usepackage{latexsym}
\usepackage{graphicx}
\usepackage{amsmath,amssymb,amsfonts}
\usepackage{bm}
\usepackage{braket}
\usepackage{color}
\usepackage{comment}
\usepackage{physics}
\usepackage[%
 setpagesize=false,%
 bookmarks=false,%
 bookmarksdepth=tocdepth,%
 bookmarksnumbered=true,%
 colorlinks=true,%
 linkcolor=red,%
 citecolor=blue,%
 urlcolor=magenta%
]{hyperref}

\newcommand{\up}{\uparrow}
\newcommand{\down}{\downarrow}

\begin{document}

%\preprint{}

\title{Metastable ferromagnetic clusters in dissipative many-body systems of polar molecules}


\author{Naoki Hara}
\email[]{hara-naoki067@g.ecc.u-tokyo.ac.up}
\author{Masaya Nakagawa}
\email[]{nakagawa@cat.phys.s.u-tokyo.ac.jp}
\affiliation{Department of Physics, University of Tokyo, 7-3-1 Hongo, Bunkyo-ku, Tokyo 113-0033, Japan}


\date{\today}

\begin{abstract}
We investigate the effect of two-body loss due to chemical reactions on quantum magnetism of fermionic polar molecules in an optical lattice. We show that an interplay between dissipation and strong long-range interactions leads to formation of metastable ferromagnetic clusters. The spin states of clusters are controlled by interaction parameters and reflect the symmetry of interactions. The size of clusters strongly depends on the initial configuration of molecules due to Hilbert-space fragmentation during dissipative many-body dynamics. We construct an effective model to show the emergence of metastable states as quasi-dark states. Application to quantum simulation of the spin-$S$ Heisenberg model is discussed.
\end{abstract}

%\keywords{}

\maketitle

\section{Introduction}\label{ref_sec_introduction}

%%%quantum magnetism with polar molecules%%%
Ultracold molecules have been of great interest in recent years since they provide a versatile platform for quantum information processing, quantum chemistry, precision metrology, and quantum simulation \cite{Baranov12, Moses16, Gadway16, Bohn17}. In particular, a unique feature of ultracold polar molecules is strong interactions between electric dipole moments, which can be manipulated by using electric fields and utilized to simulate long-range interacting many-body systems. One primary goal of quantum simulation with ultracold polar molecules is to study quantum magnetism with long-range interactions \cite{Micheli06, Barnett06, Gorshkov11_1, Gorshkov11_2, Hazzard13, Hazzard14_2, Yao18}. Experimental progress in producing molecules from cold atoms has enabled the realization of long-range interacting spin systems by encoding pseudospins in the rotational degrees of freedom of molecules \cite{Yan13, Hazzard14_1, Li22}. Furthermore, correlations between molecules have been probed with single-site resolution by using quantum-gas microscopy \cite{Rosenberg22, Christakis22}.

%%%problem of loss%%
The main obstacle for the realization of cold and dense samples of molecules is the presence of inelastic collisions due to chemical reactions, which lead to loss of molecules from the trap \cite{Ni08, Ospelkaus10, Ni10, Chotia12, Moses15, Park15, Will16, DeMarco19, Valtolina20, Son20, JRLi21}. To suppress the inelastic loss, a static electric field \cite{Avdeenkov06, Gorshkov08, Gonzalez-Martinez17, deMiranda11, Matsuda20}, microwave shielding \cite{Karman18, Schindewolf22, Bigagli23}, and the continuous quantum Zeno effect \cite{Yan13, Zhu14} have been utilized, but significant loss has still been observed for various species of molecules \cite{Christianen19, Liu20, Bause21, Gersema21, Bause23}. 

%%%dissipation engineering%%%
While dissipation is often detrimental to quantum systems, it can be exploited to prepare novel quantum many-body states that cannot be reached in thermal equilibrium \cite{Diehl08, Kraus08, Mueller12}. For example, particle losses in ultracold atomic gases have been utilized to realize metastable Mott insulators \cite{Mark12}, to control many-body dynamics \cite{Barontini13}, and to induce quantum phase transitions \cite{Tomita17}. In particular, two-body loss significantly affects the magnetic correlation and reverses its sign from antiferromagnetic to ferromagnetic in the Fermi-Hubbard systems \cite{Nakagawa20}, which has recently been observed in experiments \cite{Honda22}. The sign reversal of magnetic correlations is closely related to the formation of highly entangled Dicke states \cite{FossFeig12, Sponselee18}.
Thus, a natural question concerns whether one can use dissipative molecules to prepare interesting quantum many-body states without suppressing the unavoidable loss. However, many-body physics induced by dissipation in ultracold molecules is yet to be explored \cite{Syassen08, Zhu14, Parmee20, He20, Jamadagni21}.

%%%this work%%%
In this paper, we investigate the effect of dissipation on quantum magnetism of fermionic polar molecules trapped in an optical lattice. Here, the major distinction from the previously studied atomic systems with two-body loss \cite{FossFeig12, Nakagawa20} is that molecules possess long-range dipole-dipole interactions. We find that an interplay between the long-range interaction and dissipation due to two-body loss yields unique metastable states that cannot be seen in short-range interacting atomic systems. Specifically, while atoms form uniform ferromagnetic states in the dissipative Fermi-Hubbard systems \cite{FossFeig12, Nakagawa20}, the metastable states of dissipative polar molecules consist of ferromagnetic clusters, where the density profile of the system becomes highly inhomogeneous.

The sizes and spin states of the metastable clusters strongly reflect the initial configuration of molecules and the symmetry of the long-range spin-spin interactions. A key mechanism here is that the strong dipolar interactions severely restrict the mobility of molecules, leading to the formation of bound clusters \cite{Barbiero15, Li20}. The freezing of the motion of molecules is interpreted as effective Hilbert-space fragmentation \cite{Sala20, Khemani20} due to conservation of the number of neighboring particles \cite{Li21}. We construct an effective model that describes the dynamics of spin states of molecular clusters in each fragmented Hilbert subspace. As a result, we find that the metastable ferromagnetic clusters are described as \textit{quasi-dark states}, by which we mean eigenstates immune to dissipation in the time scale for which the effective model is valid. The lifetime of the metastable states can be estimated from a perturbation theory with respect to the inter-cluster interactions.

%%%organization of the paper%%%
The rest of this paper is organized as follows. 
In Sec.~\ref{ref_sec_model}, we introduce the model Hamiltonian and an effective Lindblad equation to describe the dissipative dynamics of ultracold molecules subject to two-body loss. 
In Sec.~\ref{ref_sec_dark}, we show that the system has ferromagnetic steady states if the interaction has SU(2) symmetry. 
In Sec.~\ref{ref_sec_simulation}, we show the formation of metastable ferromagnetic clusters from numerical simulations of the dynamics for the cases with and without the SU(2) symmetry of the interactions. 
In Sec.~\ref{ref_sec_quasidark}, we construct an effective model of molecular clusters to provide analytic understanding of the metastable states. Finally, we conclude this paper in Sec.~\ref{ref_sec_conclusion}.

\section{Model}\label{ref_sec_model}
We consider fermionic polar molecules trapped in a one-dimensional (1D) optical lattice with $L$ sites. If the optical-lattice potential is sufficiently deep, this molecular system is described by the $t$-$J$-$V$-$W$ model \cite{Gorshkov11_1,Gorshkov11_2} 
\begin{align}
H =& -t\sum_{j=0}^{L-2}\sum_{\sigma=\up,\down}(c_{j\sigma}^{\dagger}c_{j+1\sigma}+\mathrm{H.c.})\nonumber\\
&+\frac{1}{2}\sum_{i\neq j}\frac{1}{|i-j|^{3}}\biggl[J_{z}S_{i}^{z}S_{j}^{z}+\frac{J_{\perp}}{2}(S_{i}^{+}S_{j}^{-}+S_{i}^{-}S_{j}^{+})\nonumber\\
&+Vn_{i}n_{j}+W(n_{i}S_{j}^{z}+n_{j}S_{i}^{z}))\biggr],
\label{ref_eq_hamiltonian}
\end{align}
where $c_{j\sigma}$ ($ c_{j\sigma}^{\dagger} $) annihilates (creates) a molecule with spin state $\sigma$ at site $j$, and $n_{j} = c_{j\up}^{\dagger}c_{j\up} + c_{j\down}^{\dagger}c_{j\down}$ denotes the number-density operator. 
The spin operators $\bm{S}_j=(S^x_j,S^y_j,S^z_j)$ are defined by $S^x_j=(S^+_j+S^-_j)/2$, $S^y_j=(S^+_j-S^-_j)/(2i)$, and $S_{j}^{z} = (c_{j\up}^{\dagger}c_{j\up} - c_{j\down}^{\dagger}c_{j\down})/2$, where $S_{j}^{+}=c_{j\up}^{\dagger}c_{j\down}$ and $S_{j}^{-}=c_{j\down}^{\dagger}c_{j\up}$ are the ladder operators. The (pseudo)spin states $\up$ and $\down$ correspond to rotational states of molecules \cite{Gorshkov11_1, Gorshkov11_2, Yan13}. The first line of the right-hand side in Eq.~\eqref{ref_eq_hamiltonian} describes hopping of molecules between neighboring sites, and the second and third lines represent dipole-dipole interaction between molecules. The coefficients $J_{z}, J_{\perp}, V, W$ determine the strength of spin-spin interaction, spin-flip interaction, density-density interaction, and spin-density interaction, respectively. Specifically, they are related to the matrix elements of dipole moment operators and can be controlled by an external electric field \cite{Gorshkov11_1}. We note that the Hamiltonian conserves the total magnetization $\sum_jS_j^z$.

Now we include the effect of loss of molecules due to two-body inelastic collisions. When molecules collide with each other and undergo exothermic chemical reactions, a large amount of internal energy is converted to kinetic energy. Thus, molecules after an inelastic collision are quickly lost from the optical lattice. Here we follow Ref.~\cite{Zhu14} to derive an effective Lindblad equation that describes the dissipative dynamics of molecules. We assume that chemical reactions occur with a rate $\Gamma_{0}>0$ when two molecules with opposite spins occupy the same site. If the loss rate $\Gamma_{0}$ and the energy cost $E_0$ of creating a doubly occupied site are much larger than the hopping rate $t$, we can adiabatically eliminate doubly occupied sites by the second-order perturbation theory and obtain an effective Lindblad equation (with $\hbar=1$) \cite{Zhu14}
\begin{align}
\frac{d\rho}{d\tau}=\mathcal{L}\rho=-i(H_{\mathrm{eff}}\rho-\rho H_{\mathrm{eff}}^{\dagger})+\sum_{j}A_{j}\rho A_{j}^{\dagger},
\label{ref_eq_Lindblad}
\end{align}
with
\begin{equation}
H_{\mathrm{eff}}=H-\frac{i}{2}\sum_{j}A_{j}^{\dagger}A_{j}
\label{ref_eq_effective}
\end{equation}
and
\begin{align}
A_{j}=\sqrt{2\Gamma_{\mathrm{eff}}}\bigl[(c_{j\up}c_{j+1\down}&-c_{j\down}c_{j+1\up})\nonumber\\
&+(c_{j\up}c_{j-1\down}-c_{j\down}c_{j-1\up})\bigr],
\label{ref_eq_dissipator}
\end{align}
where $\rho(\tau)$ is the density matrix of the system at time $\tau$ and $\Gamma_{\mathrm{eff}}=\Gamma_0 t^2/(E_0^2+\Gamma_0^2)$ is an effective loss rate. The Lindblad operator $A_{j}$ induces a loss of a nearest-neighbor spin-singlet pair. In Eq.~\eqref{ref_eq_Lindblad}, the generator of the time evolution (the Liouvillian superoperator) is denoted by $\mathcal{L}$.

\section{Dark states}\label{ref_sec_dark}
First, we show that the system has dark states if the interactions have SU(2) spin-rotation symmetry. A dark state $\ket{D}$ is defined by \cite{Diehl08, Kraus08}
\begin{gather}
H\ket{D}=E_D\ket{D},\label{ref_eq_dark1}\\
A_j\ket{D}=0\ (\forall j).\label{ref_eq_dark2}
\end{gather}
Equations \eqref{ref_eq_dark1} and \eqref{ref_eq_dark2} lead to $\mathcal{L}(\ket{D}\bra{D})=0$, which means that a dark state is a steady state of the Lindblad equation \eqref{ref_eq_Lindblad}. 
Let $\ket{\Psi_0}$ be a spin-polarized energy eigenstate of the Hamiltonian \eqref{ref_eq_hamiltonian} that contains $N$ spin-up particles and no spin-down particle. 
This state $\ket{\Psi_0}$ is a dark state since it satisfies $A_j\ket{\Psi_0}=0$. 
Now we assume that the Hamiltonian possesses SU(2) symmetry, i.e., $J_\perp=J_z$ and $W=0$ in Eq.~\eqref{ref_eq_hamiltonian}. 
In this case, since $[S^-,H]=0$ and $[S^-,A_j]=0$ for $S^-\equiv\sum_jS_j^-$, a state
\begin{equation}
\ket{\Psi_n}\equiv\frac{1}{\sqrt{\mathcal{N}_n}}(S^-)^n\ket{\Psi_0}
\label{ref_eq_darkFM}
\end{equation}
is also a dark state of the system for any positive integer $n$ that satisfies $n\leq N$. 
Here, $\mathcal{N}_n$ denotes the normalization factor that ensures $\langle \Psi_n | \Psi_n \rangle = 1$. 
The dark state \eqref{ref_eq_darkFM} is a ferromagnetic state and fully symmetric with respect to the exchange of the spin states of the particles.
This symmetry of the many-body wave function explains the absence of loss in the dark state, since the Fermi statistics dictates that the real-space wave function be fully antisymmetric and thus forbids doubly occupied sites that cause two-body loss \cite{FossFeig12, Nakagawa20}. 
We note that the ferromagnetic dark state is analogous to the one observed in a dissipative Fermi-Hubbard system subject to two-body loss \cite{FossFeig12, Nakagawa20, Nakagawa21}.

\section{Simulation results}\label{ref_sec_simulation}
Next, we numerically calculate the dynamics of molecules subject to two-body loss. 
In the numerical simulation, we consider four molecules in a 1D lattice with $L=12$ sites under the open boundary condition. 
We assume that two of the molecules are spin up and the others are spin down. 
We employ the quantum-trajectory description of the dynamics under the Lindblad equation \eqref{ref_eq_Lindblad} \cite{Daley14} and follow the dynamics without a loss event to find stable and metastable states. 
In each quantum trajectory, the dynamics before a loss event is described by the Schr\"{o}dinger equation
\begin{equation}
i\frac{d}{d \tau}|\psi(\tau)\rangle=H_{\mathrm{eff}}|\psi(\tau)\rangle
\label{ref_eq_Schroedinger}
\end{equation}
with the effective non-Hermitian Hamiltonian \eqref{ref_eq_effective}. 
The squared norm of the state $|\psi(\tau)\rangle$ gives the probability of having no loss event until time $\tau$.
We numerically solve the Schr\"{o}dinger equation \eqref{ref_eq_Schroedinger} by using exact diagonalization of $H_{\mathrm{eff}}$. We consider two initial states
\begin{equation}
|\bm{u}\rangle=|\circ\circ\up\down\circ\circ\circ\circ\up\down\circ\circ\rangle
\label{ref_eq_u}
\end{equation}
and
\begin{equation}
|\bm{v}\rangle=|\circ\circ\up\up\circ\circ\circ\circ\down\down\circ\circ\rangle,
\label{ref_eq_v}
\end{equation}
where $\up\ (\down)$ represents the position of a spin-up (spin-down) particle and $\circ$ denotes an empty site. 
In all numerical simulations, we set the effective loss rate as $\Gamma_{\mathrm{eff}}/t = 0.1$. 
In the following, we set $t=1$ as the unit of energy so that the unit of time is given by the hopping time $\tau_0 \equiv 1/t$.

\subsection{$t$-$J$ chain\label{ref_subsec_SU(2)}}
We first consider an SU(2) symmetric case by setting $J_{z}=50$, $J_{\perp}=50$, $V=0$, and $W=0$. 
This model corresponds to the $t$-$J$ chain with long-range interactions, for which the realization with cold molecules can achieve the strongly interacting regime \cite{Gorshkov11_2}. 
Below we highlight the role of the long-range interactions that are absent in the dissipative Fermi-Hubbard model with two-body loss studied in Refs.~\cite{FossFeig12, Nakagawa20}.

\begin{figure*}
\includegraphics[width=2\columnwidth]{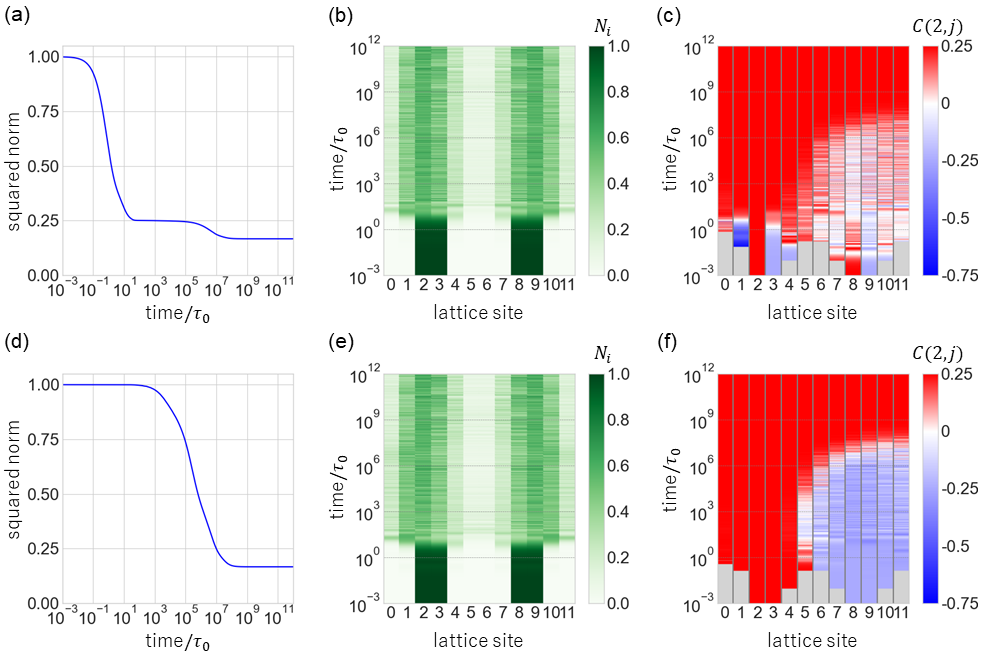}
\caption{Simulation results of the dynamics with SU(2) symmetry. The parameters of the Hamiltonian are set to $J_{z}=50$, $J_{\perp}=50$, $V=0$, and $W=0$. The initial state is $|\bm{u}\rangle$ in (a)-(c) and $|\bm{v}\rangle$ in (d)-(f). (a),(d) Time evolution of the squared norm. (b),(e) Number density $N_i$ at each site. (c),(f) Conditional spin correlation $C(2,j)$ defined in Eq.~\eqref{ref_eq_correlation}. Red (blue) color shows ferromagnetic (antiferromagnetic) correlations. When ${\langle\psi(\tau)|P(2,j)|\psi(\tau)\rangle}<0.0001$, correlations are not calculated and colored gray.\label{ref_fig_su2}}
\end{figure*}

Figure~\ref{ref_fig_su2} shows the time evolution of the squared norm $\langle \psi(\tau) | \psi(\tau) \rangle$, the number density 
\begin{equation}
N_i(\tau) \equiv \frac{\langle \psi(\tau) | {P}(i) |\psi(\tau)\rangle}{\langle \psi(\tau) | \psi(\tau) \rangle}
\label{ref_eq_density}
\end{equation}
at each site, and the conditional spin correlation 
\begin{equation}
C(i,j;\tau)\equiv\frac{\langle\psi(\tau)|P(i,j)\bm{S}_{i}\cdot\bm{S}_{j}|\psi(\tau)\rangle}{\langle\psi(\tau)|P(i,j)|\psi(\tau)\rangle}
\label{ref_eq_correlation}
\end{equation}
calculated for the initial states $|\bm{u}\rangle$ and $|\bm{v}\rangle$. 
Here $P(\cdot)$ is the projection operator onto the subspace in which all the sites in the parentheses are occupied. 
The conditional spin correlation removes the effect of motion of particles and efficiently extracts the information of spin states. 
Experimentally, it can be directly measured with the recently developed quantum-gas microscopy \cite{Nakagawa20, Hilker17,Rosenberg22, Christakis22}.
In Fig.~\ref{ref_fig_su2}, we fix $i=2$ and vary $j$ to measure the conditional correlation \eqref{ref_eq_correlation}.

Figure~\ref{ref_fig_su2}(a) shows that the squared norm for the initial state $|\bm{u}\rangle$ decreases to 0.25 at $\tau\approx 1/\Gamma_{\mathrm{eff}}$, stays constant until $\tau\approx 10^4$, and converges to 1/6 at $\tau \approx 10^8$. 
Figure~\ref{ref_fig_su2}(d) shows that the squared norm for the initial state $|\bm{v}\rangle$ is almost constant until $\tau\approx 10^{2}$, and gradually decreases to 1/6. 
The convergence to the stationary value 1/6 in both cases indicates the formation of the dark state, which is immune to dissipation.
In fact, this value of the residual squared norm is equal to the weight of the dark state in the initial states $|\bm{u}\rangle$ and $|\bm{v}\rangle$. To see this, we consider the state $\ket{\Psi_n}$ [see Eq.~\eqref{ref_eq_darkFM}] with $n=2$ and take the spin-polarized state $\ket{\Psi_0}$ as
\begin{align}
\ket{\Psi_0}=|\circ\circ\up\up\circ\circ\circ\circ\up\up\circ\circ\rangle,
\label{ref_eq_darkPsi0}
\end{align}
which can be expanded by spin-polarized eigenstates of the Hamiltonian. 
Since this state $\ket{\Psi_2}$ contains all spin configurations having the same magnetization with equal weights, we obtain 
\begin{equation}
|\langle \Psi_2 | \bm{u} \rangle|^2 = |\langle \Psi_2 | \bm{v} \rangle|^2 = \frac{1}{\binom{4}{2}} = \frac{1}{6}.
\end{equation}
The existence of the dark state is similar to the case of short-range interacting atomic systems subject to two-body loss in Refs.~\cite{FossFeig12, Nakagawa20}. 
In contrast, the plateaus at $\langle\psi(\tau)|\psi(\tau)\rangle=0.25$ and $\langle\psi(\tau)|\psi(\tau)\rangle=1$ are unique to the long-range interacting molecular system. 
These non-decaying behaviors of the squared norm indicate the formation of metastable states that defy losses during the time interval.

\begin{figure}
\includegraphics[width=\columnwidth]{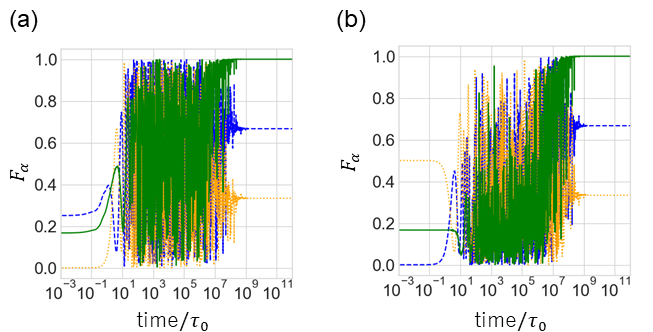}
\caption{Conditional fidelities calculated for three states, $|\bm{e}_{1}\rangle$ (blue dashed), $|\bm{e}_{2}\rangle$ (orange dotted), and $|\bm{e}_{3}\rangle$ (green solid). The initial state is $\ket{\bm{u}}$ in (a) and $\ket{\bm{v}}$ in (b). The interaction parameters are the same as those in Fig.~\ref{ref_fig_su2}. \label{ref_fig_fidelity}}
\end{figure}

The number density in Figs.~\ref{ref_fig_su2}(b) and \ref{ref_fig_su2}(e) shows that the molecules do not spread over the lattice even after the hopping time $\tau\approx 1$ and stay near their initial positions. 
This behavior also indicates that neighboring molecules form a stable group. 
We refer to these stable groups of molecules as \textit{clusters}. 
The molecular clusters persist in the dark state, which makes a sharp contrast with the atomic system where the particle distribution is uniform in the dark state \cite{Nakagawa20, Nakagawa21}.
Thus, the formation of molecular clusters is closely related to the long-range interactions.

In Figs. \ref{ref_fig_su2}(c) and \ref{ref_fig_su2}(f), the conditional spin correlations $C(i,j;\tau)$ with the molecule at site $i=2$ are plotted. 
Both of these plots show that ferromagnetic correlations within a cluster are established by $\tau\approx 1/\Gamma_{\mathrm{eff}}$, but ferromagnetic correlations between clusters are not developed until the formation of the dark state. 
The eventual development of ferromagnetic correlations throughout the system is observed in the atomic systems as well \cite{Nakagawa20}. 
However, the formation of the clusters drastically changes the transient dynamics of the spin correlations; the intercluster correlations are suppressed for a long time.

In order to clarify the nature of the spin states and their dynamics, we calculate the fidelity. 
In Fig.~\ref{ref_fig_fidelity}, the following three conditional fidelities
\begin{equation}
F_{\alpha} \equiv \frac{|\langle \bm{e}_{\alpha} | \psi(\tau) \rangle|^{2}}{\langle\psi(\tau)|P(2,3,8,9)|\psi(\tau)\rangle}
\label{ref_eq_fidelity}
\end{equation}
with $\alpha=1,2,3$ are plotted. Here,
\begin{align}
|\bm{e}_{1}\rangle=&\frac{1}{2}(\ket{\up\down}_L\ket{\up\down}_R+\ket{\up\down}_L\ket{\down\up}_R\notag\\
&+\ket{\down\up}_L\ket{\up\down}_R+\ket{\down\up}_L\ket{\down\up}_R),
\label{ref_eq_fidelity1}
\end{align}
\begin{equation}
|\bm{e}_{2}\rangle=\frac{1}{\sqrt{2}}(\ket{\up\up}_L\ket{\down\down}_R+\ket{\down\down}_L\ket{\up\up}_R),
\label{ref_eq_fidelity2}
\end{equation}
and
\begin{align}
|\bm{e}_{3}\rangle=&\frac{1}{\sqrt{6}}(\ket{\up\up}_L\ket{\down\down}_R+\ket{\down\down}_L\ket{\up\up}_R+\ket{\up\down}_L\ket{\up\down}_R\nonumber\\
&+\ket{\up\down}_L\ket{\down\up}_R+\ket{\down\up}_L\ket{\up\down}_R+\ket{\down\up}_L\ket{\down\up}_R),
\label{ref_eq_fidelity3}
\end{align}
where the kets with subscript $L$ ($R$) denote spin states at sites $j=2,3$ ($j=8,9$). 
The result shows that the state starts to change at $\tau\approx 1$ and finally reaches the state $\ket{\bm{e}_3}$. 
The steady state $\ket{\bm{e}_3}$ is fully symmetric with respect to the exchange of particles and agrees with the dark state \eqref{ref_eq_darkFM} with Eq.~\eqref{ref_eq_darkPsi0}. 
We confirmed that similar results are obtained for other choices of four sites in Eqs.~(\ref{ref_eq_fidelity})-(\ref{ref_eq_fidelity3}), in which two belong to one cluster and the other two belong to the other cluster. 
We note that the fidelities show oscillations in the region where the squared norm shows a plateau. 
It indicates that the system effectively undergoes a unitary dynamics in the metastable state. 
The underlying physics of this metastable state will be discussed in Sec.~\ref{ref_sec_quasidark}.

\subsection{$t$-$J_{\perp}$-$V$ chain}\label{ref_subsec_tjperp}
\begin{figure}
\includegraphics[width=\columnwidth]{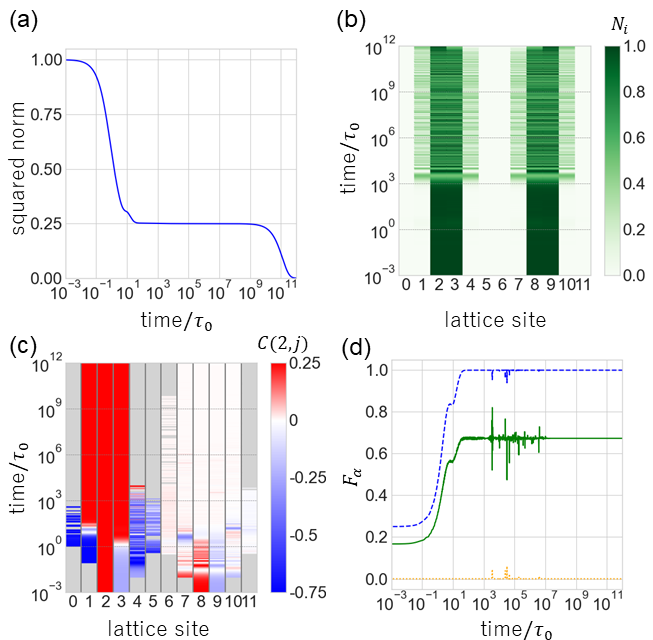}
\caption{Simulation results of the dynamics of the $t$-$J_\perp$-$V$ chain for the initial state $|\bm{u}\rangle$. The parameters are set to $J_{z}=0$, $J_{\perp}=50$, $V=50$, and $W=0$. (a) Time evolution of the squared norm. (b) Number density $N_i$ at each site. (c) Conditional spin correlation $C(2,j)$. Red (blue) color shows ferromagnetic (antiferromagnetic) correlations. When ${\langle\psi(\tau)|P(2,j)|\psi(\tau)\rangle}<0.0001$, correlations are not calculated and colored gray. (d) Conditional fidelities for $|\bm{e}_{1}\rangle$ (blue dashed), $|\bm{e}_{2}\rangle$ (orange dotted), and $|\bm{e}_{3}\rangle$ (green solid) evaluated at sites $2,3,8,9$.\label{ref_fig_tjperp_u}}
\end{figure}

Next, we relax the SU(2) symmetry of the model by setting $J_{z}=0$, $J_{\perp}=50$, $V=50$, and $W=0$. 
The model with these parameters corresponds to the $t$-$J_\perp$-$V$ chain \cite{Manmana17}.
The results for the initial state $|\bm{u}\rangle$ are shown in Fig.~\ref{ref_fig_tjperp_u}. 
Figure \ref{ref_fig_tjperp_u}(a) shows that the squared norm decreases to 0.25 in the time scale of $1/\Gamma_{\mathrm{eff}}$, stays constant until $\tau\approx 10^{9}$, and vanishes at $\tau\approx 10^{12}$. 
Thus, in contrast to the SU(2)-symmetric case, the dark state is no longer formed. 
However, a metastable state is still formed.

From Fig.~\ref{ref_fig_tjperp_u}(b), we see that the molecules are stuck to the initial positions until $\tau\approx 10^{3}$, and the density fluctuations are smaller than those in Figs.~\ref{ref_fig_su2}(b) and \ref{ref_fig_su2}(e). 
This indicates that the clusters are more robust than the case with SU(2) symmetry. 
The robustness is due to the density-density interaction $V$, as is clarified later (see Sec.~\ref{ref_sec_effmodel}).

Figure \ref{ref_fig_tjperp_u}(c) shows that the spin correlations within a cluster become ferromagnetic and saturate the maximum value at $\tau\approx 1/\Gamma_{\mathrm{eff}}$, and no correlation between clusters is developed. This result reveals the magnetic property of the metastable state. The same conclusion holds for the correlations between other sites such as $C(1,2)$ and $C(3,6)$.

The nature of the metastable state can be extracted from the conditional fidelity shown in Fig.~\ref{ref_fig_tjperp_u}(d).
The result clearly shows that the metastable state is expressed by the state $|\bm{e}_{1}\rangle$. 
Since $|\bm{e}_{1}\rangle$ can be expressed as a tensor product state of each cluster as
\begin{equation}
\ket{\bm{e}_{1}} = \frac{1}{\sqrt{2}}(\ket{\up\down}_{L}+\ket{\down\up}_{L}) \otimes \frac{1}{\sqrt{2}}(\ket{\up\down}_{R}+\ket{\down\up}_{R}),
\end{equation}
the numerically obtained spin correlations are reproduced. 
Also, the squared norm of the metastable state is explained by the squared overlap $|\langle\bm{e}_{1}|\bm{u}\rangle|^{2}=0.25$ with the initial state.

\begin{figure}
\includegraphics[width=\columnwidth]{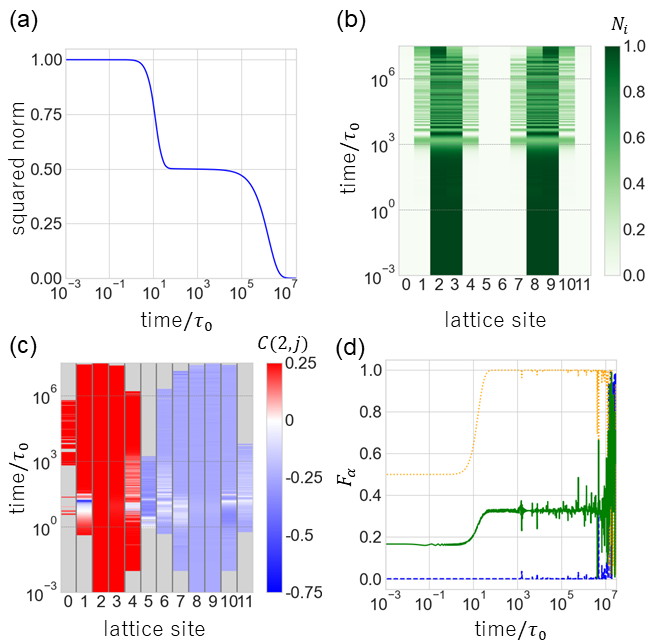}
\caption{Simulation results of the dynamics of the $t$-$J_\perp$-$V$ chain for the initial state $|\bm{v}\rangle$. The parameters are set to $J_{z}=0$, $J_{\perp}=50$, $V=50$, and $W=0$. (a) Time evolution of the squared norm. (b) Number density $N_i$ at each site. (c) Conditional spin correlation $C(2,j)$. Red (blue) color shows ferromagnetic (antiferromagnetic) correlations. When ${\langle\psi(\tau)|P(2,j)|\psi(\tau)\rangle}<0.0001$, correlations are not calculated and colored gray. (d) Conditional fidelities for $|\bm{e}_{1}\rangle$ (blue dashed), $|\bm{e}_{2}\rangle$ (orange dotted), and $|\bm{e}_{3}\rangle$ (green solid) evaluated at sites $2,3,8,9$.\label{ref_fig_tjperp_v}}
\end{figure}

The results for the initial state $|\bm{v}\rangle$ are shown in Fig.~\ref{ref_fig_tjperp_v}. Figure \ref{ref_fig_tjperp_v}(a) shows that the squared norm is almost unchanged until $\tau\approx 1$, decreases to 0.5 around $\tau\approx 1/\Gamma_{\mathrm{eff}}$, stays constant until $\tau\approx 10^{4}$, and vanishes at $\tau\approx 10^{7}$. Thus, similarly to the case for the initial state $|\bm{u}\rangle$, a metastable state is formed while the dark state is not.

The number density shown in Fig.~\ref{ref_fig_tjperp_v}(b) indicates the formation of clusters similarly to the previous cases. Figure \ref{ref_fig_tjperp_v}(c) shows that the spin correlations within a cluster are ferromagnetic, and the spin correlations between clusters stay around $-0.25$ with small fluctuations. Thus, the ferromagnetic clusters formed in this metastable state are correlated. These intercluster correlations result from spin states of the clusters which can be understood by referring to the conditional fidelity shown in Fig.~\ref{ref_fig_tjperp_v}(d). It indicates that the metastable state corresponds to the state $|\bm{e}_{2}\rangle$, in which the two clusters are entangled. The value of the squared norm at the plateau is consistent with the squared overlap $|\langle\bm{e}_2|\bm{v}\rangle|^2=0.5$.


Another useful indicator of the existence of metastable states is the spectrum of the non-Hermitian Hamiltonian \eqref{ref_eq_effective}. In fact, since the Lindblad operator \eqref{ref_eq_dissipator} decreases the particle number by two, the Liouvillian $\mathcal{L}$ defined in Eq.~\eqref{ref_eq_Lindblad} can be represented as a triangular matrix in a particular basis, and its eigenvalues are obtained from the diagonal entries which are determined solely from the non-Hermitian Hamiltonian \cite{Torres14, Nakagawa21}. Specifically, an eigenvalue $\lambda_{ab}$ of the Liouvillian $\mathcal{L}$ is given by a pair of eigenvalues $E_a$ and $E_b$ of the non-Hermitian Hamiltonian $H_{\mathrm{eff}}$ as $\lambda_{ab}=-i(E_a-E_b^*)$. Since the absolute value of the real part of an eigenvalue $\lambda_{ab}$ corresponds to the decay rate of the corresponding eigenmode of the Liouvillian, eigenvalues with small absolute values of the imaginary part in the spectrum of the non-Hermitian Hamiltonian indicate the existence of metastable states with a long lifetime.

\begin{figure}
\includegraphics[width=\columnwidth]{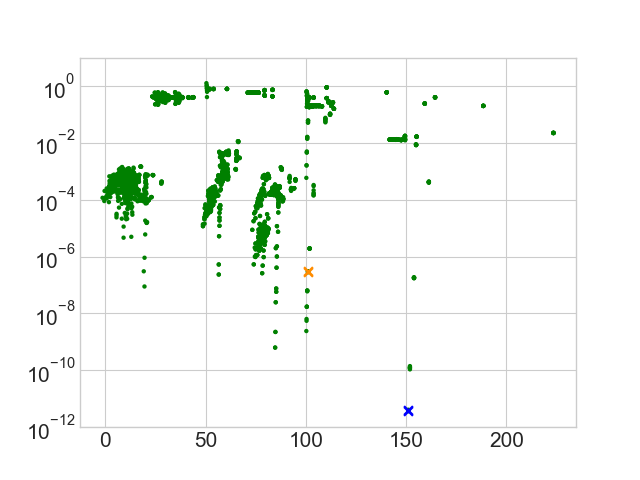}
\caption{Spectrum of the effective non-Hermitian Hamiltonian $H_{\mathrm{eff}}$ with $J_{z}=0$, $J_{\perp}=50$, $V=50$, and $W=0$. The horizontal axis shows the real part of eigenvalues, and the vertical axis shows the absolute value of the imaginary part of eigenvalues. A blue (orange) cross corresponds to the metastable state for the initial state $|\bm{u}\rangle$ ($|\bm{v}\rangle$).\label{ref_fig_tjperp_spectrum}}
\end{figure}

The spectrum of the effective non-Hermitian Hamiltonian \eqref{ref_eq_effective} is plotted in Fig.~\ref{ref_fig_tjperp_spectrum}. It contains eigenvalues with small absolute values of the imaginary part, each of which corresponds to a metastable state with different configuration of particles. An eigenvalue corresponding to the metastable state for the initial state $|\bm{u}\rangle$ ($|\bm{v}\rangle$) is marked with a blue (orange) cross. The inverse of the imaginary part of the eigenvalues shows a good agreement with the lifetime of the metastable states in Figs.~\ref{ref_fig_tjperp_u} and \ref{ref_fig_tjperp_v}. 
We note that the spectrum contains several eigenvalues with nonzero absolute values of the imaginary part smaller than $10^{-12}$, which are not shown in Fig.~\ref{ref_fig_tjperp_spectrum}. 
They correspond to metastable states in which the clusters are separated by a distance longer than that in the initial states considered here. 
We confirmed that their spin states are similar to those of the metastable states in the numerical simulation.

\subsection{Larger-size clusters}\label{ref_subsec_largersize}
\begin{figure}
\includegraphics[width=\columnwidth]{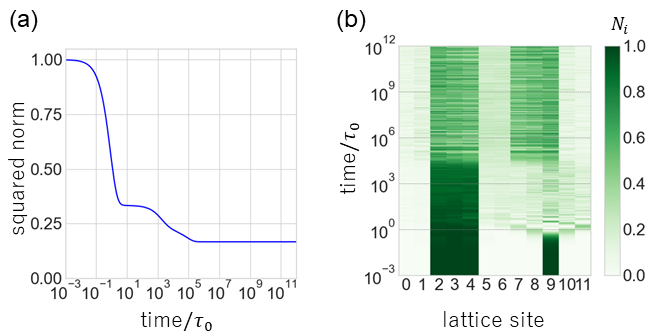}
\caption{Simulation results for the initial state $\ket{\bm{w}}$. The parameters are set to $J_{z}=50$, $J_{\perp}=50$, $V=0$, and $W=0$. (a) Time evolution of the squared norm. (b) Number density $N_i$ at each site.\label{ref_fig_su2_size3}}
\end{figure}

In the previous subsections, we have shown the formation of clusters of two molecules. To investigate clusters with larger sizes, we calculate the dynamics with the initial state
\begin{equation}
|\bm{w}\rangle=|\circ\circ\up\down\up\circ\circ\circ\circ\down\circ\circ\rangle.
\label{ref_eq_w}
\end{equation}
Figure \ref{ref_fig_su2_size3} shows the results for the SU(2) symmetric Hamiltonian with $J_z=50,J_\perp=50,V=0$, and $W=0$. The squared norm in Fig.~\ref{ref_fig_su2_size3}(a) decreases to 1/3 at $\tau\approx 1/\Gamma_{\mathrm{eff}}$, stays constant until $\tau\approx 10^{2}$, and converges to 1/6 at $\tau\approx 10^6$. The plateau of the squared norm at the value 1/3 signals the formation of a metastable state, and the stationary value of the squared norm indicates that the system reaches the dark state as in Sec.~\ref{ref_subsec_SU(2)}. The number density in Fig.~\ref{ref_fig_su2_size3}(b) shows that the metastable state consists of a size-three cluster and a delocalized molecule, while the dark state is expressed by the superposition of two clusters. 

\begin{figure}
\includegraphics[width=\columnwidth]{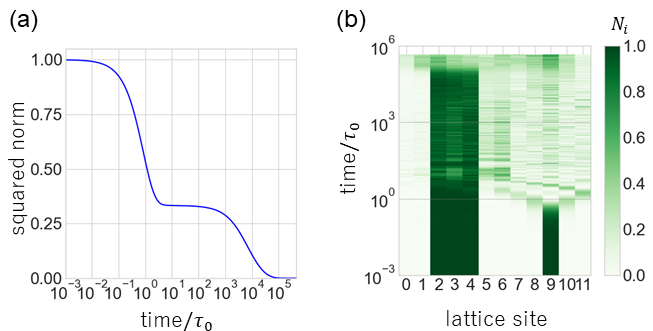}
\caption{Simulation results for the initial state $\ket{\bm{w}}$. The parameters are set to $J_{z}=80$, $J_{\perp}=100$, $V=1.25$, and $W=10$. (a) Time evolution of the squared norm. (b) Number density $N_i$ at each site.\label{ref_fig_size3}}
\end{figure}

We also find that a metastable cluster of three molecules can be formed even if the interactions lack SU(2) symmetry. Here, we employ the parameters $J_{z}=80, J_{\perp}=100, V=1.25$, and $W=10.$ The time evolution of the squared norm shown in Fig.~\ref{ref_fig_size3}(a) indicates that a metastable state is formed while dark state is not, although the lifetime of the metastable state is much shorter than that of the previous cases. The number density shown in Fig.~\ref{ref_fig_size3}(b) indicates that the metastable state consists of a size-three cluster and a delocalized molecule. We confirmed that the spin state of the metastable cluster is given by the W state \cite{Dur00}
\begin{equation}
\frac{1}{\sqrt{3}}(\ket{\up\up\down} + \ket{\up\down\up} + \ket{\down\up\up}).
\label{ref_eq_3meta}
\end{equation}


\section{Effective model for metastable states}\label{ref_sec_quasidark}
\subsection{Effective model and quasi-dark states}\label{ref_sec_effmodel}
Here we construct a simple, effective model to explain the emergence of the metastable ferromagnetic states in the long-range interacting dissipative fermionic molecular system. As found numerically, molecules in clusters are almost localized at their initial positions. Similar behavior has been observed in dipolar bosons on a lattice without dissipation \cite{Barbiero15, Li20, Korbmacher22, Aramthottil22}. The mechanism behind the formation of clusters becomes clear if one takes the $V\to\infty$ limit, where the numbers $\langle n_{j}n_{j+l}\rangle\ (l=1,2,\cdots)$ of molecular pairs at distance $l$ are conserved. 
% revised sentence (C)
If the strength of the long-range interaction is finite, $\ev{n_{j}n_{j+l}}$ is approximately conserved up to a certain $l$ in a time scale that scales exponentially with the ratio between the energy cost to violate this conservation and the hopping amplitude \cite{Li21,Strohmaier10,Sensarma10}.
The localization of particles in long-range interacting systems has recently been interpreted in terms of Hilbert-space fragmentation \cite{Sala20, Khemani20, Li21}, where the dynamics of the system is confined in a Hilbert subspace due to the kinetic constraint imposed by strong long-range interactions. Hence, to simplify the discussion, we put an assumption that the positions of molecules are fixed to their initial ones and construct an effective model for the spin degrees of freedom of molecules.

Here we note a difference between spinless bosons in the previous work \cite{Li21} and spinful fermions in the present work. 
%In fact, metastable clusters in Sec.~\ref{ref_subsec_SU(2)} are formed without the density-density interaction $V$. 
The numerical calculation in Sec.~\ref{ref_subsec_SU(2)} shows that the metastable clusters are formed even without the density-density interaction $V$.
This implies that the strong spin-spin interaction suffices for the formation of clusters as the (approximate) conservation of $\langle S_j^+S_{j+l}^-+S_j^-S_{j+l}^+\rangle$ and $\langle S_j^zS_{j+l}^z\rangle$ ($l=1,2,\cdots$) can lead to Hilbert-space fragmentation. If one additionally has a strong density-density interaction, the clusters become more stable. This effect can be seen from the suppression of density fluctuations shown in Sec.~\ref{ref_subsec_tjperp}.

To derive an effective model that describes spin states of molecular clusters, we decompose the spin interaction Hamiltonian $H_{\mathrm{spin}}$ as
\begin{align}
H_{\mathrm{spin}}\equiv&\frac{1}{2}\sum_{i\neq j}\frac{1}{|i-j|^{3}}\biggl[J_{z}S_{i}^{z}S_{j}^{z}+\frac{J_{\perp}}{2}(S_{i}^{+}S_{j}^{-}+S_{i}^{-}S_{j}^{+})\nonumber\\
&+W(n_{i}S_{j}^{z}+n_{j}S_{i}^{z}))\biggr]\nonumber\\
=&H_{\mathrm{intra}}+H_{\mathrm{inter}},
\end{align}
where $H_{\mathrm{intra}}$ ($H_{\mathrm{inter}}$) denotes the interaction term acting on molecules that belong to the same (different) cluster(s).
If the distance between clusters is sufficiently long, the intra-cluster interactions are dominant. Thus, we first diagonalize the intra-cluster interaction term as
\begin{equation}
H_{\mathrm{intra}}=\sum_nE_nP_n,
\end{equation}
and treat the inter-cluster interactions as perturbation. Here, $P_n$ denotes the projection operator onto the eigenspace $\mathcal{H}_n$ corresponding to eigenvalue $E_n$. Then, the effective model within an eigenspace $\mathcal{H}_n$ is given by
\begin{equation}
H_{\mathrm{eff}}^{(n)}=E_nP_n+P_nH_{\mathrm{inter}}P_n-\frac{i}{2}\sum_{j}P_nA_{j}^{\dagger}A_{j}P_n,
\label{ref_eq_effective_n}
\end{equation}
where the last term corresponds to the anti-Hermitian term due to dissipation in Eq.~\eqref{ref_eq_effective}.

Given the effective model, we define a \textit{quasi-dark state} $\ket{\tilde{D}}\in\mathcal{H}_n$ as an eigenstate of $H_{\mathrm{eff}}^{(n)}$ that is annihilated by the Lindblad operators:
\begin{align}
H_{\mathrm{eff}}^{(n)}\ket{\tilde{D}}=E_{\tilde{D}}\ket{\tilde{D}},\label{ref_eq_qdark1}\\
A_j\ket{\tilde{D}}=0\ (\forall j).\label{ref_eq_qdark2}
\end{align}
A quasi-dark state describes a metastable state by construction since it is immune to dissipation up to a time scale for which the effective model is valid.

\subsection{Comparison with numerical results}
The simplest case of metastable states is given by two clusters, each of which is formed by two molecules as shown in the numerical simulation in Sec.~\ref{ref_sec_simulation}. We fix the locations of four molecules at $i, i+1, j, j+1$, and denote the spin state of the cluster at sites $i$ and $i+1$ ($j$ and $j+1$) by kets with subscript $L$ ($R$). In this case, eigenvalues $E_n$ and eigenspaces $\mathcal{H}_n$ of the intra-cluster interaction term are given by
\begin{align}
E_1=&\frac{J_z}{2},\ %\notag\\
\mathcal{H}_1=\mathrm{span}[\{\ket{\up\up}_{L}\ket{\down\down}_{R},\ket{\down\down}_{L}\ket{\up\up}_{R}\}],\label{ref_eq_E1}\\
E_2=&J_\perp-\frac{J_z}{2},\ %\notag\\
\mathcal{H}_2=\mathrm{span}[\{\ket{t}_{L}\ket{t}_{R}\}],\label{ref_eq_E2}\\
E_3=&-\frac{J_z}{2},\ %\notag\\
\mathcal{H}_3=\mathrm{span}[\{\ket{t}_{L}\ket{s}_{R},\ket{s}_{L}\ket{t}_{R}\}],\label{ref_eq_E3}\\
E_4=&-J_\perp-\frac{J_z}{2},\ %\notag\\
\mathcal{H}_4=\mathrm{span}[\{\ket{s}_{L}\ket{s}_{R}\}],\label{ref_eq_E4}
\end{align}
where each ket represents a spin state of a cluster, $\ket{t}\equiv\frac{1}{\sqrt{2}}(\ket{\up\down}+\ket{\down\up})$, and $\ket{s}\equiv\frac{1}{\sqrt{2}}(\ket{\up\down}-\ket{\down\up})$. For simplicity, here we set $W=0$ and consider the states with zero total magnetization. We have also assumed that the eigenvalues $E_1,E_2,E_3,E_4$ are not degenerate.

The Lindblad operator $A_j$ induces a loss of a spin-singlet pair in neighboring sites. Namely, we have $A_j\ket{s}=\sqrt{2\Gamma_{\mathrm{eff}}}\ket{\circ\circ}$ and $A_j\ket{\up\up}=A_j\ket{\down\down}=A_j\ket{t}=0$, where $j$ denotes a site occupied by a molecule in a cluster. Thus, if the eigenvalues $E_1,E_2,E_3,E_4$ are not degenerate, all eigenstates of the effective Hamiltonians $H_{\mathrm{eff}}^{(1)}$ and $H_{\mathrm{eff}}^{(2)}$ are quasi-dark states, while states in the other eigenspaces $\mathcal{H}_3$ and $\mathcal{H}_4$ are not. 
% newly added sentence (A)
Since the lifetime of the singlet state $\ket{s}$ is given by the inverse of the two-body loss rate $2\Gamma_{\mathrm{eff}}$, the timescale of the formation of metastable states is given by $1/(2\Gamma_{\mathrm{eff}})$, which agrees with the numerical results in Sec.~\ref{ref_sec_simulation}.

\subsubsection{$t$-$J$ chain}
The metastable states observed in the numerical simulation in Sec.~\ref{ref_sec_simulation} are understood from the effective model. Let us consider the SU(2)-symmetric case (i.e., the $t$-$J$ chain) analyzed in Sec.~\ref{ref_subsec_SU(2)}. Then, the eigenvalues $E_1$ and $E_2$ in Eqs.~\eqref{ref_eq_E1} and \eqref{ref_eq_E2} are degenerate since $J_\perp=J_z$. Therefore, while all the eigenstates in the eigenspace $\mathcal{H}_1\oplus\mathcal{H}_2$ become quasi-dark states, they are mixed by the action of the inter-cluster interaction $H_{\mathrm{inter}}$ and the system undergoes an effective unitary dynamics in the space of quasi-dark states.

For the initial state $\ket{\bm{u}}$, the formation of a metastable state is indicated by the plateau of the squared norm at the value $0.25$ [see Fig.~\ref{ref_fig_su2}(a)]. This value is understood from the overlaps between quasi-dark states and the initial state; in fact, we have $|({}_{L}\bra{t}{}_{R}\bra{t})\ket{\bm{u}}|^2=0.25$ and $|({}_{L}\bra{\up\up}{}_{R}\bra{\down\down})\ket{\bm{u}}|^2=|({}_{L}\bra{\down\down}{}_{R}\bra{\up\up})\ket{\bm{u}}|^2=0$ for the sites with $i = 2$ and $j = 8$. The plateau for the initial state $\ket{\bm{v}}$ is understood similarly since this initial state already belongs to the eigenspace $\mathcal{H}_1$. The dynamics in the space of quasi-dark states is inferred from the numerical results of the spin correlations and the fidelities (see Figs.~\ref{ref_fig_su2} and \ref{ref_fig_fidelity}).

The lifetime of the metastable states can be estimated from a perturbation theory. The actions of the inter-cluster interactions on the states $\ket{P}\equiv\frac{1}{\sqrt{2}}(\ket{\up\up}_{L}\ket{\down\down}_{R}+\ket{\down\down}_{L}\ket{\up\up}_{R})$, $\ket{Q}\equiv\frac{1}{\sqrt{2}}(\ket{\up\up}_{L}\ket{\down\down}_{R}-\ket{\down\down}_{L}\ket{\up\up}_{R})$, and $\ket{t}_L\ket{t}_R$ are given by
\begin{align}
H_{\mathrm{inter}}&\ket{P}\notag\\
&= -\frac{J_z}{4}\qty[\frac{2}{r^3}+\frac{1}{(r+1)^3}+\frac{1}{(r-1)^3}]\ket{P}\notag\\
&\ \ + \frac{J_\perp}{2\sqrt{2}}\qty[\frac{2}{r^3}+\frac{1}{(r+1)^3}+\frac{1}{(r-1)^3}]\ket{t}_{L}\ket{t}_{R}\notag\\
&\ \ - \frac{J_\perp}{2\sqrt{2}}\qty[\frac{2}{r^3}-\frac{1}{(r+1)^3}-\frac{1}{(r-1)^3}]\ket{s}_{L}\ket{s}_{R},\label{ref_eq_interp}
\end{align}
\begin{align}
H_{\mathrm{inter}}&\ket{Q}\notag\\
&= -\frac{J_z}{4}\qty[\frac{2}{r^3} + \frac{1}{(r+1)^3} + \frac{1}{(r-1)^3}]\ket{Q}\notag\\
&\ \ - \frac{J_\perp}{2}\qty[\frac{1}{(r+1)^3}-\frac{1}{(r-1)^3}]\frac{1}{\sqrt{2}}(\ket{t}_{L}\ket{s}_{R}+\ket{s}_{L}\ket{t}_{R}),\label{ref_eq_interq}
\end{align}
\begin{align}
H_{\mathrm{inter}}&\ket{t}_{L}\ket{t}_{R}\notag\\
&= \frac{J_\perp}{2\sqrt{2}}\qty[\frac{2}{r^3}+\frac{1}{(r+1)^3}+\frac{1}{(r-1)^3}]\ket{P}\notag\\
&\ \ + \frac{J_z}{4}\qty[\frac{2}{r^3}-\frac{1}{(r+1)^3}-\frac{1}{(r-1)^3}]\ket{s}_{L}\ket{s}_{R}\label{ref_eq_intert}
\end{align}
where $r=|i-j|$.
Thus, quasi-dark states are mixed with the decaying states that contain $\ket{s}$ by the first-order perturbation of the inter-cluster interaction. 
The lifetime is given by the inverse of the absolute value of the imaginary part of the eigenvalues, which can be estimated from the second-order perturbation theory.
The imaginary parts of the eigenvalues of the metastable states $\ket{P}$, $\ket{Q}$, and $\ket{t}_L\ket{t}_R$ are given by
\begin{align}
&\ket{P}: && \mathrm{Im}\qty[\frac{\qty{\frac{J_\perp}{2\sqrt{2}}\qty[\frac{2}{r^3}-\frac{1}{(r+1)^3}-\frac{1}{(r-1)^3}]}^2}{J_z + J_\perp + 2i\Gamma_{\mathrm{eff}}}],\label{ref_eq_lifeP}\\
&\ket{Q}: && \mathrm{Im}\qty[\frac{\qty{\frac{J_\perp}{2}\qty[\frac{1}{(r+1)^3}-\frac{1}{(r-1)^3}]}^2}{J_z + i\Gamma_{\mathrm{eff}}}],\label{ref_eq_lifeQ}\\
&\ket{t}_L\ket{t}_R: && \mathrm{Im}\qty[\frac{\qty{\frac{J_z}{4}\qty[\frac{2}{r^3}-\frac{1}{(r+1)^3}-\frac{1}{(r-1)^3}]}^2}{2J_\perp + 2i\Gamma_{\mathrm{eff}}}]\label{ref_eq_lifett}.
\end{align}
For the cases calculated in Sec.~\ref{ref_subsec_SU(2)}, we have $J_\perp=J_z=50t$, $\Gamma_{\mathrm{eff}}=0.1t$, and $r=6$. These parameters give the lifetime of the order of $10^8\tau_0$ for $\ket{P}$ and $\ket{t}_L\ket{t}_R$ and $10^6\tau_0$ for $\ket{Q}$. These values agree with the results of numerical simulation, which indicate that the metastable states are lost at $\tau\approx 10^8\tau_0$.
% newly added sentence (B)
It is worth noting that the lifetime of metastable states $\ket{P}$ and $\ket{t}_{L}\ket{t}_{R}$ is proportional to $r^{10}$, while that of $\ket{Q}$ scales as $r^{8}$.

\subsubsection{$t$-$J_\perp$-$V$ chain\label{ref_subsec_tJperpVmeta}}
In the case of $J_z=0$ analyzed in Sec.~\ref{ref_subsec_tjperp}, the eigenvalues $E_1$ and $E_3$ in Eqs.~\eqref{ref_eq_E1} and \eqref{ref_eq_E3} are degenerate. From Eqs.~\eqref{ref_eq_interp} and \eqref{ref_eq_interq}, a quasi-dark state in the eigenspace $\mathcal{H}_1\oplus\mathcal{H}_3$ is given by $\ket{P}$. This quasi-dark state corresponds to the metastable state realized for the initial state $\ket{\bm{v}}$ shown in Fig.~\ref{ref_fig_tjperp_v}, where the value of the squared norm at the plateau is understood from the squared overlap $|\langle P | \bm{v} \rangle|^2=0.5$ for $i = 2$ and $j = 8$. This is corroborated by the fidelity shown in Fig.~\ref{ref_fig_tjperp_v}. Here it is worthwhile to note that this quasi-dark state has quantum entanglement between clusters. Thus, we can induce inter-cluster entanglement by controlling the spin-spin interaction so that a part of states in the Hilbert subspace $\mathcal{H}_1$ is dissipated due to mixing with states in another Hilbert subspace $\mathcal{H}_3$. The lifetime of this quasi-dark state is evaluated from Eq.~\eqref{ref_eq_lifeP}, which is consistent with the numerical result in Fig.~\ref{ref_fig_tjperp_v}.

The metastable state reached from the initial state $\ket{\bm{u}}$ in Fig.~\ref{ref_fig_tjperp_u} is given by the quasi-dark state $\ket{t}_L\ket{t}_R$ in the eigenspace $\mathcal{H}_2$, which is the same as in the SU(2) symmetric case. However, we note that the lifetime of this quasi-dark state is longer than that in the SU(2)-symmetric case by orders of magnitude. This is because the imaginary part \eqref{ref_eq_lifett} of the eigenvalue of this state vanishes up to the second order of the inter-cluster interactions. The leading contribution to the imaginary part comes from a fourth-order process, which can be approximately estimated as
\begin{align}
\mathrm{Im}\left[\frac{(J_\perp/r^3)^4}{(J_\perp+2i\Gamma_{\mathrm{eff}})J_\perp^2}\right]\sim-\frac{2\Gamma_{\mathrm{eff}}}{r^{12}}.
\end{align}
Thus, the lifetime of the metastable state in this case is of the order of $10^{10}\tau_0$ for $\Gamma_{\mathrm{eff}}=0.1t$ and $r=6$. This estimate agrees with the time evolution of the squared norm shown in Fig.~\ref{ref_fig_tjperp_u}.

\subsection{Extension to general clusters}

Now we discuss the general structure of metastable states of the system on the basis of the effective model. 
Suppose that the system has $M$ clusters with size $m_1,m_2,\cdots,m_M$. Here, a cluster with size $m$ consists of $m$ neighboring molecules and is stabilized by the approximate conservation of $\langle n_{j}n_{j+l}\rangle$, $\langle S_j^+S_{j+l}^-+S_j^-S_{j+l}^+\rangle$, or $\langle S_j^zS_{j+l}^z\rangle$ due to strong long-range interactions, as discussed in Sec.~\ref{ref_sec_effmodel}. 
To find a metastable state, we first diagonalize the intra-cluster interaction Hamiltonian and identify non-dissipative states that are annihilated by the Lindblad operators. If the interaction is SU(2) symmetric, the solution is given by a tensor product of $m_a$-particle ferromagnetic states [similar to Eq.~\eqref{ref_eq_darkFM}] with spin $m_a/2$ ($a=1,\cdots,M$) since it is a dark state of the intra-cluster Hamiltonian (see Sec.~\ref{ref_sec_dark}). For $m_a=2$, the states $\ket{\up\up},\ket{\down\down}$, and $\ket{t}$ generally give the non-dissipative states as shown in Eqs.~\eqref{ref_eq_E1}-\eqref{ref_eq_E4}. If $m_a\geq3$ and the interaction is not SU(2) symmetric, such non-dissipative states do not exist in general, but we can find them if the interaction strengths are fine-tuned. An example is given by Eq.~\eqref{ref_eq_3meta} in Sec.~\ref{ref_subsec_largersize}. The W state \eqref{ref_eq_3meta} is an eigenstate of the intra-cluster Hamiltonian if the interaction strengths satisfy $J_\perp=J_z+2W$, which leads to $[S_{\mathrm{intra}}^{-},H_{\mathrm{intra}}]\ket{\uparrow\uparrow\uparrow}=0$ with $S_{\mathrm{intra}}^-\equiv\sum_{j\in \mathrm{cluster}}S^-_{j}$.
% newly added sentence
%In addition, strong long-range interaction is required to suppress the decay of clusters due to approaches of other clusters or singlons, since there are resonant hoppings like $\ket{\up \down \up \circ \down} \leftrightarrow \ket{\up\down\circ\up\down}$ which conserve the number of nearest-neighboring pairs. This kind of hoppings can be suppressed if the number of next-nearest-neighboring pairs are conserved, and it is realized when the energy cost paid to violate this conservation is larger than the bandwidth $4t$ \cite{Li21}.

After the intra-cluster interaction Hamiltonian is diagonalized, we consider the effect of the inter-cluster interactions. If a state in an eigenspace $\mathcal{H}_n$ of the intra-cluster interaction Hamiltonian is not coupled to a dissipative state within the eigenspace, it describes a metastable state. For the SU(2)-symmetric case, it is convenient to introduce a spin operator $\bm{T}_a$ for the $a$th cluster ($a=1,\cdots,M$) as
\begin{equation}
\bm{T}_a=\sum_{j=j_a}^{j_a+m_a-1}\bm{S}_j,
\end{equation}
where the positions of molecules in the $a$th cluster are denoted by $j_a,j_a+1,\cdots,j_a+m_a-1$. An intra-cluster ferromagnetic state is an eigenstate of $\bm{T}_a^2$ with eigenvalue $(m_a/2)(m_a/2+1)$. If the inter-cluster interaction Hamiltonian is SU(2) symmetric, it also commutes with $\sum_{a}\bm{T}_a$, and thus the effective model up to the lowest order of the inter-cluster interactions should be written as
\begin{equation}
H_{\mathrm{eff}}^{(n)}=\sum_{a,b=1}^M K_{a,b}\bm{T}_a\cdot\bm{T}_b,
\label{ref_eq_THeis}
\end{equation}
where $K_{a,b}$ is a coupling constant. Thus, the metastable ferromagnetic clusters undergo an effective unitary dynamics under the Heisenberg Hamiltonian \eqref{ref_eq_THeis}.

Finally, we write down general metastable states for the case of $m_1=m_2=\cdots=m_M=2$ and when the interaction is not SU(2) symmetric. In this case, a tensor-product state $\ket{t}^{\otimes M}$ of the state $\ket{t}$ of each cluster becomes metastable because of Eq.~\eqref{ref_eq_intert}. To write down more nontrivial metastable states, we regard the states $\ket{\Uparrow}_a\equiv\ket{\up\up}_a$ and $\ket{\Downarrow}_a\equiv\ket{\down\down}_a$ as pseudospin states, where the subscript $a$ denotes the $a$th cluster. The pseudospin operators are defined by
\begin{align}
\tilde{S}_a^x=&\frac{1}{2}(\ket{\Uparrow}_a\bra{\Downarrow}_a+\ket{\Downarrow}_a\bra{\Uparrow}_a),\\
\tilde{S}_a^y=&\frac{1}{2i}(\ket{\Uparrow}_a\bra{\Downarrow}_a-\ket{\Downarrow}_a\bra{\Uparrow}_a),\\
\tilde{S}_a^z=&\frac{1}{2}(\ket{\Uparrow}_a\bra{\Uparrow}_a-\ket{\Downarrow}_a\bra{\Downarrow}_a),
\end{align}
and the ladder operators are given by $\tilde{S}_a^\pm=\tilde{S}_a^x\pm i\tilde{S}_a^y$. Using the pseudospin operators, we define a cluster-symmetric state as
\begin{equation}
\ket{\tilde{\Psi}_m}\equiv\frac{1}{\sqrt{\mathcal{N}_m}}(\tilde{S}^-)^m\left[\bigotimes_{a=1}^M\ket{\Uparrow}_a\right],
\end{equation}
where $\mathcal{N}_m$ is a normalization factor, $\tilde{S}^-\equiv \sum_{a=1}^M\tilde{S}_a^-$, and $m\leq M$. Analogous to the dark state \eqref{ref_eq_darkFM}, this state $\ket{\tilde{\Psi}_m}$ is symmetric with respect to exchange of the pseudospin states of clusters. Therefore, this state can be written as
\begin{align}
\ket{\tilde{\Psi}_m}=&\frac{1}{\sqrt{2}}(\ket{\up\up}_a\ket{\down\down}_b+\ket{\down\down}_a\ket{\up\up}_b)\otimes\ket{\phi_1}\notag\\
&+\ket{\up\up}_a\ket{\up\up}_b\otimes\ket{\phi_2}+\ket{\down\down}_a\ket{\down\down}_b\otimes\ket{\phi_3},
\end{align}
where $\ket{\phi_1},\ket{\phi_2}$, and $\ket{\phi_3}$ denote states of the clusters except for the $a$th and $b$th ones. 
If $J_z=0$, the states $\frac{1}{\sqrt{2}}(\ket{\up\up}_a\ket{\down\down}_b+\ket{\down\down}_a\ket{\up\up}_b)$, $\ket{\up\up}_a\ket{\up\up}_b$, and $\ket{\down\down}_a\ket{\down\down}_b$ are not coupled to the energetically degenerate decaying states $\ket{t}_a\ket{s}_b$ and $\ket{s}_a\ket{t}_b$ by the inter-cluster interactions because of Eqs.~\eqref{ref_eq_interp} and \eqref{ref_eq_interq}. Thus, the state $\ket{\tilde{\Psi}_m}$ becomes metastable. 
This generalizes the metastable state $\ket{\bm{e}_2}$ [Eq.~\eqref{ref_eq_fidelity2}] with quantum entanglement between clusters that is observed for the case of $M=2$ in Sec.~\ref{ref_subsec_tJperpVmeta}.

\section{Conclusion}\label{ref_sec_conclusion}
We have shown that ultracold polar molecules with strong long-range interactions form metastable states in which ferromagnetic clusters avoid inelastic losses for a considerably longer time than the inverse of the two-body loss rate. The metastable states significantly depend on the initial condition, which is a consequence of ergodicity breaking due to Hilbert-space fragmentation caused by the approximate conservation of the numbers of neighboring pairs. Since the quasi-dark metastable states cannot be formed in short-range interacting atomic systems \cite{Nakagawa20, Honda22, Sponselee18}, this result implies that non-ergodic dynamics in the Hamiltonian part can be exploited to enrich dissipative quantum many-body dynamics. We note that even if the Hilbert-space fragmentation is incomplete due to the finiteness of the interaction strengths, the thermalization time scale is anomalously enhanced in such strongly interacting systems because relaxation of an energetically costly state requires exponentially long times \cite{Strohmaier10, Sensarma10}. Thus, even if the loss rate is suppressed by some means such as the continuous quantum Zeno effect \cite{Yan13, Zhu14}, inelastic collisions may have a significant impact on quantum magnetism of polar molecules in a long time scale.

Some of our results can be extended to bosonic molecules. In the case of two-component bosons, two-body loss tends to develop antiferromagnetic correlations \cite{Nakagawa20}. However, if the number of bosons is larger than two, the system does not have a dark state since it requires a fully antisymmetric many-body spin wavefunction, which does not exist in general except for the two-body spin-singlet state $\ket{s}$. Thus, a possible metastable state for bosonic pola	r molecules is a tensor-product state $\ket{s}^{\otimes M}$ of $M$ clusters of spin singlets.

% newly added sentences
Another possible extension of our result is metastable states in higher dimensions. The cluster dynamics in two dimensions is studied in Refs.~\cite{Salerno20,Li21cl}, where clusters are similarly formed due to long-range interactions and perform a quantum walk reflecting a lattice geometry. On the other hand, clusters with sizes larger than three become more fragile than those in one dimension because of the presence of various resonant processes that break up clusters. Whereas the fate of the Hilbert space fragmentation of molecular systems in higher dimensions requires further investigation, our effective model for metastable states is applicable to the case of higher dimensions as long as clusters of molecules are stabilized by long-range interactions.

Finally, we note that the system in the metastable state undergoes an effective unitary dynamics described by the Heisenberg Hamiltonian \eqref{ref_eq_THeis} if the spin-spin interaction has SU(2) symmetry. This feature may be exploited for quantum simulation of the spin-$S$ Heisenberg model with ultracold polar molecules in the presence of loss. The magnitude $S_a$ of the spin $\bm{T}_a$ of the $a$th cluster is given by the size $m_a$ of the cluster as $S_a=m_a/2$, which is controlled by the initial configuration of molecules. Our results thus not only show the unique many-body dynamics of dissipative polar molecules, but also open a way for their application to quantum simulation.

\begin{acknowledgments}
We are grateful to Masahito Ueda for helpful discussions. 
M.N. was supported by KAKENHI Grant No.~JP20K14383 from the Japan Society for the Promotion of Science. 
\end{acknowledgments}

\bibliography{reference.bib}

\end{document}